# DEKGCI: A double-sided recommendation model for integrating knowledge graph and user-item interaction graph

Yajing Yang, Zeyu Zeng, Mao Chen*, Ruirui Shang

*Faculty of Artificial Intelligence in Education, Central China Normal University, Wuhan 430079, China*

*Corresponding author. E-mail address: chenmao@ccnu.edu.cn (M. Chen)

**Abstract**

Both knowledge graphs and user-item interaction graphs are frequently used in recommender systems due to their ability to provide rich information for modeling users and items. However, existing studies often focused on one of these sources (either the knowledge graph or the user-item interaction graph), resulting in underutilization of the benefits that can be obtained by integrating both sources of information. In this paper, we propose DEKGCI, a novel double-sided recommendation model. In DEKGCI, we use the high-order collaborative signals from the user-item interaction graph to enrich the user representations on the user side. Additionally, we utilize the high-order structural and semantic information from the knowledge graph to enrich the item representations on the item side. DEKGCI simultaneously learns the user and item representations to effectively capture the joint interactions between users and items. Three real-world datasets are adopted in the experiments to evaluate DEKGCI's performance, and experimental results demonstrate its high effectiveness compared to seven state-of-the-art baselines in terms of AUC and ACC.

## 1. Introduction

The advent of the Internet era has led to an overwhelming explosion of information. Coping with the vast amount of data can be time-consuming and overwhelming, leading to decreased efficiency in resource utilization and decision-making when input data exceeds one's processing capabilities. To address this issue of information overload, information filtering systems have been developed to eliminate redundant or unnecessary information and provide personalized recommendations to users based on their interests. One type of information filtering system is recommender systems, which aim to predict a user's preference or rating for an item. These systems have been successfully applied in various domains such as news websites (Huang et al., 2022), social media platforms (Liao et al., 2021; Boeker & Urman, 2022), online shopping (Lu et al., 2013; Linden et al., 2003), and other web applications (Hu et al., 2020; Khalid et al., 2022; Mcsherry & Mironov, 2009) over the past few decades, providing great convenience to consumers and enhancing the competitiveness and profitability of these businesses (Andrea et al., 2022).

There are two main approaches to building recommender systems, namely content-based and collaborative filtering (Verbert et al., 2012; Mooney & Roy, 2000; Breese et al., 2013; Balabanović & Shoham, 1997). Unlike the content-based approach that requires knowledge of the features of users and items, collaborative filtering calculates similarities between users based on their historical behaviors and recommends items that similar users have liked to a target user. However, despite its widespread use in various fields, collaborative filtering still faces challenges such as data sparsity and cold start problems (Wang et al., 2017; Zhu et al., 2019). To address these issues, researchers often leverage side information such as user attributes (Cheng et al., 2016; Wang et al., 2018a), item attributes (Zhen et al., 2009), social networks (Jamali & Ester, 2010), and images (Zheng et al., 2016) to compensate for data scarcity and improve recommendation performance.

As a rich source of semantic information, knowledge graphs (KGs) have gained significant attention in recent years (Huang et al., 2018; Wang et al., 2018b; Wang et al., 2018c; Yu et al., 2014; Zhang et al., 2016; Zhao et al., 2017), and numerous models have been proposed to integrate KGs into recommender systems. For example, the well-known KGAT model (Wang et al., 2019a) constructs a collaborative knowledge graph (CKG) by fusing the user-item interaction graph and the KG, and propagates this information through embedding of its neighbors to generate a target representation. The KCAN model (Tu et al., 2021) refines the KG into a local subgraph using attention mechanism to obtain more target-specific information. MANN (Wu et al., 2022), a multi-context-aware KG-based recommendation algorithm, learns item representations by combining path-based and propagation-based approaches. KGs have been shown to capture user features and long-distance relationships more accurately between users and items, making it easier to discover hidden user-item connections and improve recommendation accuracy, diversity, and interpretability, as discussed in (Liu et al., 2019; Zhang et al., 2019).

However, current research on KG-based recommender systems primarily focuses on the item side, using KGs to obtain more accurate vector representations of items, while neglecting the user side of the recommender systems. Despite some recent attempts (Li et al., 2022; Sun et al., 2022), existing double-sided recommendation methods still lack the ability to learn deeper information about user-item interactions to improve user representations (Wang et al., 2019b). For example, Li et al. (2022) constructs user representations using auxiliary data such as the user's age and occupation, and obtains item representations by aggregating heterogeneous information from the KG, but the high-level user-item interaction information has not been fully explored.

Motivated by previous studies (Wang et al., 2019a; Wang et al., 2019b; He et al., 2020), this work proposes a double-sided recommendation model based on KG and high-level collaborative information (DEKGCI) to address the aforementioned shortcomings of previous studies. DEKGCI aims to obtain the embedding vector of items from the KG's structure information and semantic information, extract neighbor node's features through high-order user-item interactions to enrich the user's representation, and finally combine the item and user representations for the final recommendation. By learning both the user and item representations simultaneously, DEKGCI overcomes the limitations of single-sided modeling. Moreover, compared to previous KG-based recommendation methods and double-sided recommendation models, the proposed method can fully leverage the high-order user-item interaction information to enhance the user representation and improve the performance of recommendations. Three real-world benchmarks, namely movie, book, and music, are used to evaluate the performance of DEKGCI, and it is compared to seven state-of-the-art methods for click-through rate (CTR) prediction, including MANN (Wu et al., 2022), KGCN (Wang et al., 2019c), and RippleNet (Wang et al., 2018b). The experimental results demonstrate that DEKGCI outperforms these baselines in terms of AUC and ACC.

The following are the paper's main contributions:

- DEKGCI is a successful double-sided recommendation model, which takes into account both the user side and the item side of the recommender system. It utilizes the user-item interaction graph to enrich the user representation and the knowledge graph (KG) to enrich the item representation. Furthermore, during the model training process, the user representations and item representations are optimized simultaneously, leading to a more comprehensive and effective recommendation approach.

- Extensive experiments on three widely used datasets demonstrate that the proposed method is highly competitive compared to seven state-of-the-art algorithms. DEKGCI achieves significant improvements in terms of AUC and ACC. Specifically, it achieves AUC gains of 1.7% and 3.1%, and ACC gains of 4.8% and 2.9% for movie and book recommendations, respectively. In addition, DEKGCI also achieves a gain of 0.5% in AUC for music recommendation, further validating its effectiveness and superiority compared to existing methods in the literature.

## 2. Related work

### 2.1 User modeling

In order to make accurate recommendations, it is important to have a comprehensive understanding of users' preferences, including both static (e.g., interests) and dynamic (e.g., user-item interactions) information. Some recommender systems (Zhang et al., 2020) take into account both types of information to better model user preferences. With the integration of knowledge graphs (KGs) into recommender systems, the rich semantic information in KGs, combined with user-item interactions, are jointly utilized to model users' preferences. For instance, in (Wang et al., 2020), collaborative data extracted from user-item interactions is combined with knowledge information to obtain user vector embeddings. The KGE-CF model (Xu et al., 2021) maps previously interacted items of users to entities in the KG, and then employs a translation model to learn the vector embeddings of entities and their relationships, before learning users' embedding representations based on preference propagation.

In contrast to the above-mentioned studies that only consider direct user-item connections (i.e., one-hop neighbors), the NGCF model (Wang et al., 2019b) aims to enhance user representations by utilizing high-order connectivity. Specifically, NGCF uses neighborhood aggregation enabled by graph convolutional networks (GCN) (Kipf & Welling, 2016) to take into account the effects of multi-hop neighbors, and learns user embeddings by propagating them on the user-item bipartite graph. To further improve the effectiveness of the NGCF model, a lightweight version called LightGCN is proposed (He et al., 2020), which removes redundant components from NGCF such as feature transformation and nonlinear activation, resulting in a more streamlined and efficient approach.

**2.2 Item modeling**

The rich structure information and semantic information in KGs are utilized to enhance item representations in various ways. For instance, in the CKE model (Zhang et al., 2016), semantic representations of items, including textual and visual representations, are extracted from structural knowledge, textual knowledge, and visual knowledge, respectively. These representations are then combined with collaborative filtering to generate the latent representation of items. In the DKN model (Wang et al., 2018c), sequential training is employed to generate entity embeddings and word embeddings for news articles, which are then input into a convolutional neural network (CNN) framework as multiple channels to generate news representations. The KGCN model (Wang et al., 2019c) aggregates information from multi-hop neighbors to generate entity representations, extending the GCN model to KG-based recommendations. As shown in Fig.1, the representation of the entity *The Million Pound Note* is dependent on its 1-order neighbors, namely *Critique* and *Mark Twain*, and similarly, the representation of its 1-order neighbors is dependent on its 2-hop neighbor, the entity *The Prince and the Pauper*. In this way, the long-distance item-item relatedness are captured to model the items.

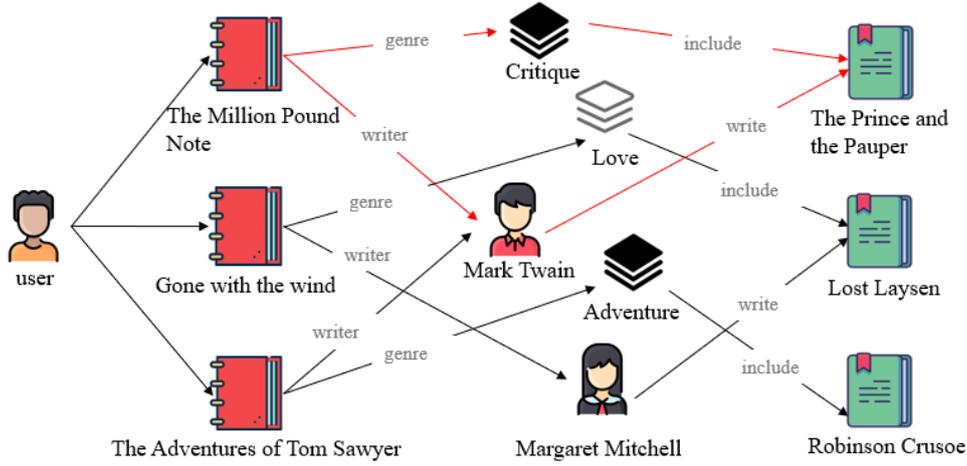

**Fig. 1.** An example of the process of embedding the entity *The Million Pounds* in a knowledge graph using information from 1-hop and 2-hop neighbors

## 3. Problem formulation

If there is an interaction (e.g., ratings, purchases) between user *u* and item *v* in a recommendation scenario with a set of users $U=\{u_1, u_2, ..., u_m\}$ and a set of items $V=\{v_1, v_2, ..., v_n\}$, then $y_{uv}$, the corresponding element in interaction matrix $Y$ is labeled as 1, otherwise, $y_{uv}=0$. The items and the relationships between them are structured as a KG, denoted as $G$, which consists of a number of triples $(h, r, t)$, where $h$ represents the head entity, $r$ represents the relationship, and $t$ represents the tail entity.

The goal of this study is to learn the embedding representation of the user *u* and the item *v* and predict whether *u* is potentially interested in *v* through the prediction function $\hat{y}_{uv} = f(u, v|\Theta, Y, G)$, where $\Theta$ represents the model parameters.

## 4. Methodology

The overall framework of DEKGCI is depicted in Fig.2, which mainly consists of three layers: the embedding propagation layer, the aggregation layer, and the prediction layer. In the embedding propagation layer, the latent collaborative signal in the user-item interaction graphs is captured to enrich the user embedding representation, as well as the high-order connectivity in KG is exploited to model the items more accurately. In the aggregation layer, the final representation of the user is created by aggregating the user's embedding representation with the embedding representation of its neighbors, and the final representation of the item is obtained in a similar way. Finally, the user representation and the item representation are combined in the prediction layer to produce the user's clicking probability for the given item.

**4.1 User Representation**

*4.1.1 Embedding Propagation Layer*

The user-item interaction history can be visualized by a user-item interaction graph, as shown in Fig.3, and the proposed model aims to extract the collaborative signal from this history. For example, from the interaction graph in Fig.3, the high-order connectivity such as $\{u_1 \rightarrow v_1 \rightarrow u_2\}$ and $\{u_1 \rightarrow v_2 \rightarrow u_3 \rightarrow v_6\}$ can be used to more accurately mode $u_1$'s embedding representation.

As we all know, items that are directly related to a user can reveal the user's preferences. Therefore, the contribution of the 1-order neighbor $v$ to the user $u$'s feature representation can be defined as:

$$I_{u,v}^{(1)} = \frac{1}{\sqrt{|N_v||N_u|}} W_1^{(1)} e_v^{(0)} \tag{1}$$

Where $N_v$ denotes the set of users that are directly interacted with the item $v$, $N_u$ denotes the set of items that are directly interacted with the user $u$, $W_1^{(1)}$ is the trainable feature transformation matrix, and $e_v^{(0)}$ is the representation of the item $v$. The norm $1/\sqrt{|N_v||N_u|}$ is a normalization processing, which can avoid the problem of increase of embedding size that GCN causes. In contrast to the complex message encoding function in NGCF model, here we consider only the contributions of $e_v^{(0)}$ and this simplifies the model to some extent without compromising its effectiveness.

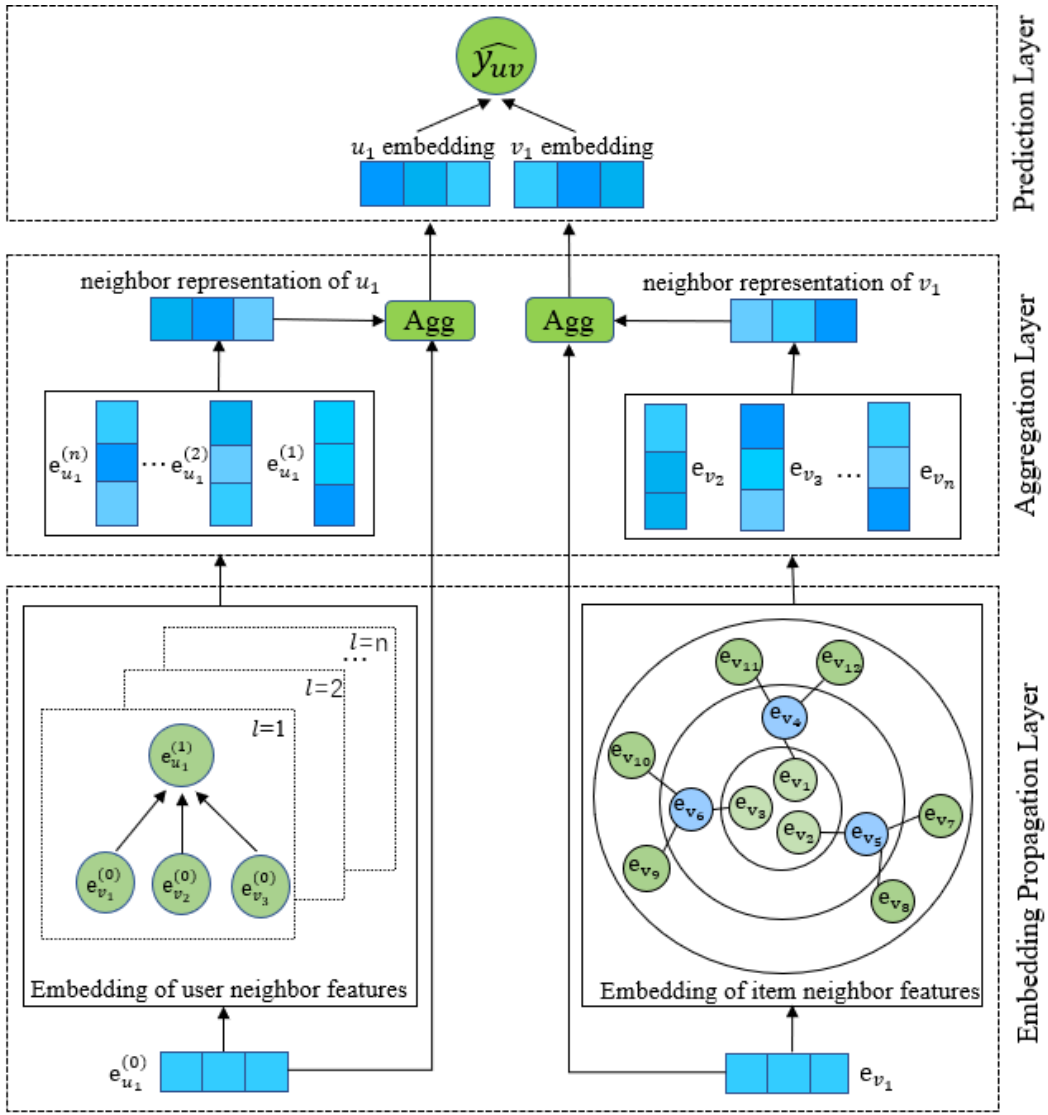

**Fig. 2.** The framework of the DEKGCI. Based on the KG and user-item interaction history, the user $u$'s click probability for the item $v$ can be calculated after three successive layers of processing.

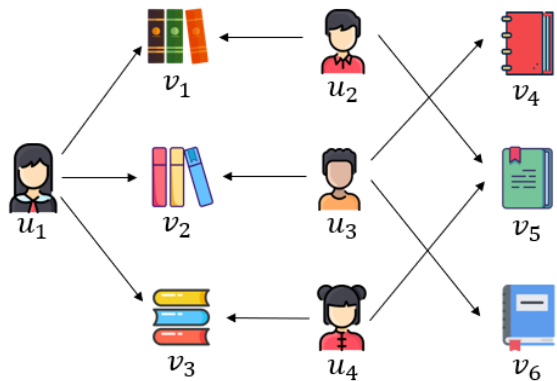

**Fig. 3.** An illustration of the high-order user-item interaction graph

In this manner, all the 1-order neighbor items' features can be aggregated to yield the user $u$'s representation according to the GCN mechanism, followed by a nonlinear transformation (i.e., the

activation function *leakyReLU*):

$$e_u^{(1)} = LeakyReLU\left(\sum_{v \in N_u} I_{u,v}^{(1)}\right) \qquad (2)$$

By stacking $l$ embedding propagation layers, a user is capable of receiving the messages propagated from its $l$-hop neighbors, in the $l$-th step, the representation of user $u$ is recursively formulated as:

$$e_u^{(l)} = LeakyReLU\left(\sum_{v \in N_u} I_{u,v}^{(l)}\right) \qquad (3)$$

wherein the messages being propagated are defined as follows:

$$I_{u,v}^{(l)} = \frac{1}{\sqrt{|N_v||N_u|}} W_1^{(l)} e_v^{(l-1)} \qquad (4)$$

*4.1.2 Aggregation layer*

After $l$ layer propagation, the embedding representation of user $u$ in layers 1 to $l$ are obtained and collected in the set $\{e_u^{(1)}, e_u^{(2)}, ..., e_u^{(l)}\}$. As the user representation produced at different layers have their own contribution to reflect the user preference, it is a natural and logical way to aggregate these embedding vectors to get the complete final representation of the user. There are mainly three aggregation methods as follows:

**Sum aggregation**. The sum aggregation sums the embedding vectors of the user at each layer up to obtain the final user representation $e_u^*$, including the initial user representation $e_u^{(0)}$:

$$e_u^* = e_u^{(0)} + e_u^{(1)} + e_u^{(2)} + ... + e_u^{(l)} \qquad (4)$$

**Concatenate aggregation**. The user representations at each layer are concatenated to obtain the final user representation $e_u^*$, including the initial user representation $e_u^{(0)}$:

$$e_u^* = e_u^{(0)} \| e_u^{(1)} \| e_u^{(2)} \| ... \| e_u^{(l)} \qquad (5)$$

Where $\|$ denotes the concatenation operator.

**Neighbor aggregation**. Without considering the initial user representation, neighbor aggregation only concatenates the user's representation vectors at layers 1 to $l$ to obtain the final user representation $e_u^*$:

$$e_u^* = e_u^{(1)} \| e_u^{(2)} \| ... \| e_u^{(l)} \qquad (6)$$

The effectiveness of the three aggregation methods will be assessed and compared in the following experiment section.

### 4.2 Item Representation

*4.2.1 Embedding Propagation layer*

Knowledge graph embedding learning aims to extend item representations by utilizing knowledge graphs' rich auxiliary information and higher-order graph structure to uncover deep features. Consider a candidate pair of user u and item (entity) v, user *u* might focus more on the subject when selecting books to read than the author. In order to discern the difference of entities connected by different relations, this model uses an attention mechanism to assess the significance of relation *r* to user *u* in the knowledge graph, which is known as the *user-relation score* (Wang et al., 2019c):

$$\hat{\pi}(u, r) = e_u^{*T} * e_r \tag{7}$$

Where $e_u^*$ and $e_r$ are the representations of users u and relationships r.

Considering all user-relation pairs, the user-relation scores can be normalized by using the Softmax function as follows:

$$\pi(u, r_{v,t}) = \frac{exp\,(\hat{\pi}(u,r_{v,t}))}{\sum_{t \in N(v)} exp(\hat{\pi}(u,r_{v,t}))} \tag{8}$$

Where $N(v)$ represents the set of item *v*'s first-order neighbors, $r_{v,t}$ represents the relationship between the item *v* and its neighbor *t*.

According to the attention mechanism, the representations of all 1-order neighbors of item *v* are multiplied by their respective normalized user-relations scores and then summed to form the first-order neighbor representation of item *v*:

$$V_{N(v)} = \sum_{t \in N(v)} \pi(u, r_{v,t}) * e_t \tag{9}$$

Where $e_t$ is the embedding representation of entity *t*.

The multi-order neighboring information in the knowledge can be aggregated layer by layer, allowing the deeper information to be extracted and the item representation to be optimized. The performance of the recommender system is sensitive to the depth of the receptive field, as shown in (Wang et al., 2019c). In this work, we only aggregate the information of the 1-order neighbors to generate the final item representation, which simplifies the model while having no negative effects on performance, as the experiments will show.

*4.2.2 Aggregation Layer*

At the Aggregation layer, the representation of the item itself is aggregated with $v$'s neighbors, i.e., $V_{N(v)}$, to generate the final item representation:

$$e_v^* = LeakyReLU(W_2*(v_0+V_{N(v)})+b) \tag{10}$$

Where $W_2$ and $b$ are the weight and bias of the transformation.

**4.3 Prediction layer**

For the click-through rate prediction, we perform inner product of user and item representations to predict the matching probability, i.e., the user's preference to the item:

$$\hat{y}_{uv} = sigmoid(e_u^{*T} * e_v^*) \tag{11}$$

The following objective function is used to learn the model parameters. To make computation more efficient, we employ a binary cross-entropy loss function:

$$L = \sum_{(u,v)\in P} -(y_{uv}*\log(\hat{y}_{uv}) + (1-y_{uv})*\log(1-\hat{y}_{uv})) \tag{12}$$

Where $P$ is the mini-batch, $y_{uv}$ is the true value, $\hat{y}_{uv}$ is the predicted value.

**5. Experiments**

**5.1 Dataset**

The proposed method is implemented in Python 3.9 with PyTorch 1.10 APIs. In order to assess and contrast the model's results, this paper runs experiments on three benchmark datasets from the movie, book, and music domains. Table 1 displays the detailed statistical results of these benchmarks. Data sparsity refers to the ratio of elements without interactive data to the whole matrix space in the user-item matrix, which is calculated as 1 minus half of the number of interactions divided by the product of the number of users and the number of items

- MovieLens-1M: A movie rating dataset with 376,887 ratings from 6,036 users on 2,445 movies and with a data sparsity of 97.45%.
- Book-Crossings: A dataset of 69,874 ratings from 17,860 users on 14,967 books, with a data sparsity of 99.97%.
- Last.FM: A music rating dataset with 21,174 hits on 3,846 singers from 1,872 users from online music websites and a data sparsity of 99.70%.

Please note that the knowledge graphs utilized in our experiments were directly sourced from the websites https://github.com/hwwang55/KGCN and https://github.com/hwwang55/RippleNet, both of which were specifically constructed to align with the aforementioned three data sets.

Table 1. Hyper-parameters settings for the three datasets

|  | MovieLens-1M | Book-Crossing | Last.FM |
|---|---|---|---|
| Users | 6036 | 17860 | 1872 |
| Items | 2445 | 14967 | 3846 |
| Interactions | 753772 | 139746 | 42346 |
| sparsity | 97.45% | 99.97% | 99.70% |
| KG entities | 182011 | 77903 | 9366 |
| KG relations | 12 | 25 | 60 |
| KG triples | 1241995 | 151500 | 15518 |

**5.2 Parameter Setting**

Table 2 displays the experimental parameter settings for the three datasets, where *batchsize* is the batch size, *N_neighbor* is the knowledge graph's sample number of neighbor nodes, and it adopts random sampling, dim is the dimension of user and item vector representation, *lr* is the learning rate, and *layer* is the number of layers in higher-order interaction graph.

All datasets are split into training set, eval set and test sets with a 6:2:2 ratio. As mentioned above, click-through rate is used to predict the user's preference to the item. Specifically, if the click-through rate > 0.5, the user will interact with the item; otherwise, the user has no interest in this item. Finally, two evaluation metrics, AUC and ACC, are used to assess the effectiveness of the model.

Table 2. Basic Parameter Settings for the three datasets

|  | MovieLens-1M | Book-Crossing | Last.FM |
|---|---|---|---|
| $batchsize$ | 1024 | 16 | 32 |
| $N\_neighbor$ | 10 | 8 | 8 |
| $dim$ | 128 | 16 | 64 |
| $lr$ | 0.0005 | 0.0005 | 0.0005 |
| $Layer$ | 3 | 6 | 3 |

**5.3 Experimental results**

*5.3.1 Computational results*

To verify the effectiveness of the DEKGCI model, the computational results are compared with the state-of-the-art methods in the literature, including MANN, KGCN, RippleNet, CKE (Zhang et al., 2016), PER (Yu et al., 2014), LibFM (Rendle, 2012), and Wide&Deep (Cheng et al., 2016).

The experimental results of DEKGCI and the reference methods are shown in Table 3. Except for MANN, DEKGCI outperforms all other state-of-the-art methods. It should be noted that MANN is a highly efficient KG-based recommendation model that combines the benefits of both path-based and propagation-based methods. In comparison to MANN, DEKGCI achieves AUC gains of 1.86%, 4.17%, and 0.60% in movie, book, and music recommendation, and ACC gains of 5.95% and 4.26% in movie and book recommendation, respectively; the result of DEKGCI is only slightly worse in ACC in music recommendation.

The experimental results and comparisons show that the four methods (i.e., MANN, KGCN, RippleNet, and DEKGCI) that use KG as a source of side information outperform the rest methods, namely CKE, PER, LibFM, and Wide&Deep, indicating the importance of side information in the recommender system. Furthermore, it can be observed from Table 3 that the overall performance of DEKGCI is the best among the four KG-based methods, confirming the key idea of our method, i.e., learning both the user representation and the item representation simultaneously in the proposed double-sided model, is effective and efficient.

**Table 3.** The results of AUC and ACC in CTR prediction

| Model | MovieLens-1M | | Book-Crossing | | Last.FM | |
|---|---|---|---|---|---|---|
| | AUC | ACC | AUC | ACC | AUC | ACC |
| MANN | 0.912 | 0.807 | 0.744 | 0.681 | 0.836 | **0.791** |
| KGCN | 0.919 | 0.848 | 0.702 | 0.640 | 0.804 | 0.731 |
| RippleNet | 0.921 | 0.844 | 0.729 | 0.668 | 0.805 | 0.735 |
| CKE | 0.910 | 0.841 | 0.677 | 0.635 | 0.744 | 0.673 |
| PER | 0.712 | 0.667 | 0.623 | 0.588 | 0.633 | 0.596 |
| LibFM | 0.892 | 0.812 | 0.691 | 0.639 | 0.777 | 0.709 |
| Wide&Deep | 0.903 | 0.822 | 0.711 | 0.623 | 0.756 | 0.688 |
| **DEKGCI** | **0.929** | **0.855** | **0.775** | **0.710** | **0.841** | 0.769 |

*5.3.2 Statistical analysis*

To see if there is a statistical difference between DEKGCI and the reference algorithm, we conduct Friedman and Iman-Davenport tests on each dataset with a significance factor of 0.05(Derrac et al., 2011). The null hypothesis assumes that there is no significant difference between the results of these algorithms. In the statistical analysis, the AUC and ACC scores are combined to provide a thorough comparison between the nine algorithms. The Friedman statistic is distributed according to chi-square with 7-degrees of freedom, while the Iman-Davenport statistic is distributed according to F-distribution with 7 and 35 degrees of freedom. Table 4 shows the results of the Friedman and Iman-Davenport tests (a = 0.05).

The null hypothesis was rejected by the Friedman and Iman-Davenport tests (p-value a), indicating a significant difference between the algorithms considered (as shown in Table 4). Following that, a post hoc Holm test is performed to determine whether or not there was a statistical difference between DEKGCI and the other reference algorithms. The post-hoc comparison results are shown in Table 5.

The statistical results in Table 5 show that DEKGCI is statistically better than PER, CKE, LibFM, and Wide&Deep. In terms of AUC and ACC, DEKGCI outperforms almost all other algorithms; however, the differences between DEKGCI and KGCN, RippleNet, and MANN are not statistically significant. The reason for this could be that the insufficient number of datasets used in our tests reduces the statistical analysis's sensitivity.

**Table 4.** Results of the Friedman and Iman-Davenport tests (a = 0.05).

| Friedman value | Value in $\chi^2$ | $p$-value | Iman-Davenport value | Value in $F_F$ | $p$-value |
|---|---|---|---|---|---|
| 34.0555556 | 14.06714045 | <0.0001 | 21.43356658 | 2.285235173 | <0.0001 |

**Table 5.** Results of the Post Hoc comparisons.

| i | Algorithm | $z=(R_0-R_i)/SE$ | $p$ | Holm |
|---|---|---|---|---|
| 1 | PER | 4.831896 | 0.000001 | 0.007143 |
| 2 | CKE | 3.417683 | 0.000632 | 0.008333 |
| 3 | LibFM | 3.181981 | 0.001463 | 0.01 |
| 4 | Wide&Deep | 3.181981 | 0.001463 | 0.0125 |
| 5 | KGCN | 1.767767 | 0.0771 | 0.016667 |
| 6 | MANN | 1.296362 | 0.194851 | 0.025 |
| 7 | RippleNet | 1.178511 | 0.238593 | 0.05 |

### 5.4 Analysis and discussion

*5.4.1 Influence of embedding propagation method*

Embedding propagation is one of the key ingredients to learn the user representation. NGCF and Light GCN, like our work, use embedding propagation to capture the collaborative filtering signal from user-item interaction graph to enrich the user representation. Specifically, NGCF adopts the following recursive formula:

$$e_u^{(l)} = \sigma(W_1^{(l)} e_u^{(l-1)} + \sum_{v \in N_u} \frac{1}{\sqrt{|N_v||N_u|}} (W_1^{(l)} e_u^{(l-1)} + W_2^{(l)}(e_v^{(l-1)} \odot e_u^{(l-1)}))) \quad (13)$$

Where $e_u^{(l)}$ denotes the refined embedding of user $u$ after $l$ layers of propagation, σ is the activation function, $N_v$ is the set of 1-order neighbors of item $v$, $N_u$ is the set of first-order neighbors of user $u$, $W_1$ and $W_2$ are trainable transformation matrices. Conversely, LightGCN greatly simplifies the above model by removing the activation function and the feature transformation matrix, which is as follows:

$$e_u^{(l)} = \sum_{v \in N_u} \frac{1}{\sqrt{|N_v||N_u|}} e_v^{(l-1)} \tag{14}$$

Apparently, only the neighbor information is used to produce the node's representation in LightGCN, whereas the information about the user itself is not included. We replace the embedding propagation approach used by DEKGCI with those used by NGCF and LightGCN, respectively. The resulting methods are referred to as DEKGCI-1 and DEKGCI-2, respectively, and they are compared with DEKGCI in terms of AUC and ACC on the three datasets to assess the effectiveness of the embedding propagation approach in DEKGCI. Table 4 displays the experimental findings.

**Table 6.** Comparison of the effect of embedding propagation methods

| Model | MovieLens-1M | | Book-Crossing | | Last.FM | |
|---|---|---|---|---|---|---|
| | AUC | ACC | AUC | ACC | AUC | ACC |
| DEKGCI | 0.929 | 0.855 | **0.775** | **0.710** | **0.841** | **0.769** |
| DEKGCI-1 | 0.915 | 0.838 | 0.755 | 0.694 | 0.809 | 0.733 |
| DEKGCI-2 | **0.933** | **0.859** | 0.743 | 0.672 | 0.837 | 0.757 |

As shown in Table 6, DEKGCI obtains better results than DEKGCI-1 and DEKGCI-2 in most case, demonstrating the value of keeping the feature transformations and activation functions in the embedding propagation method. In general, our strategy falls somewhere between NGCF and LightGCN in terms of the model complexity, but it is more efficient than that of NGCF and LightGCN.

*5.4.2 Influence of number of propagation layers*

The high-order collaborative information is valuable for enhancing the user representation. However, as pointed out in (Wang et al., 2019b), the long-distance user-item interaction will also bring noises into the representation learning, causing the model to overfit. Therefore, we conduct additional experiments with layers ranging from 1 to 6 to investigate how model performance varies with the number of embedding propagation layers in this section.

Fig. 4 show the effects of different layers on DEKGCI for the three datasets. Generally speaking,

it can be observed from Fig.4(a)~(c) that the DEKGCI's performance increases as the number of layers increases from 1 to 3; however, when the number of layers continues to increase, the performance drops for the MovieLens-1M and Last.FM datasets. After evaluating the algorithm's performance across these three datasets as a whole, we can see that three layers is the ideal number in MovieLens-1M and Last.FM datasets and six layers is the ideal number in Book-Crossing datasets.

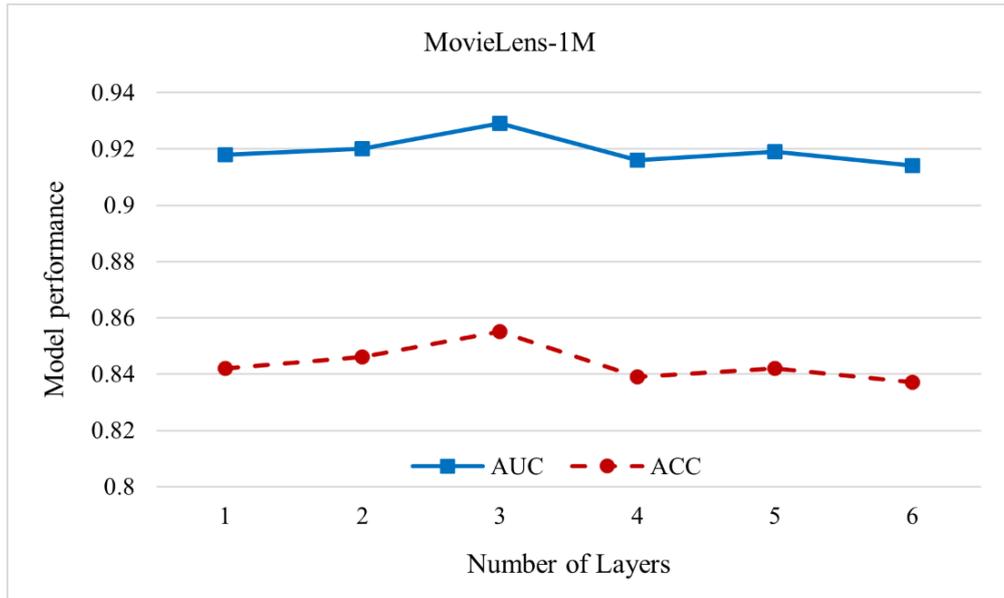

(a)

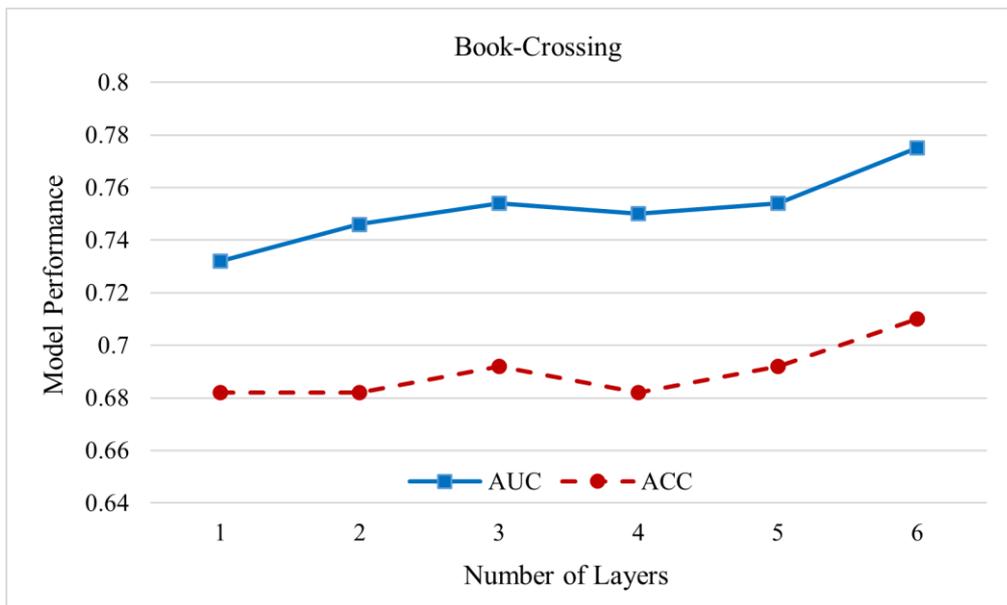

(b)

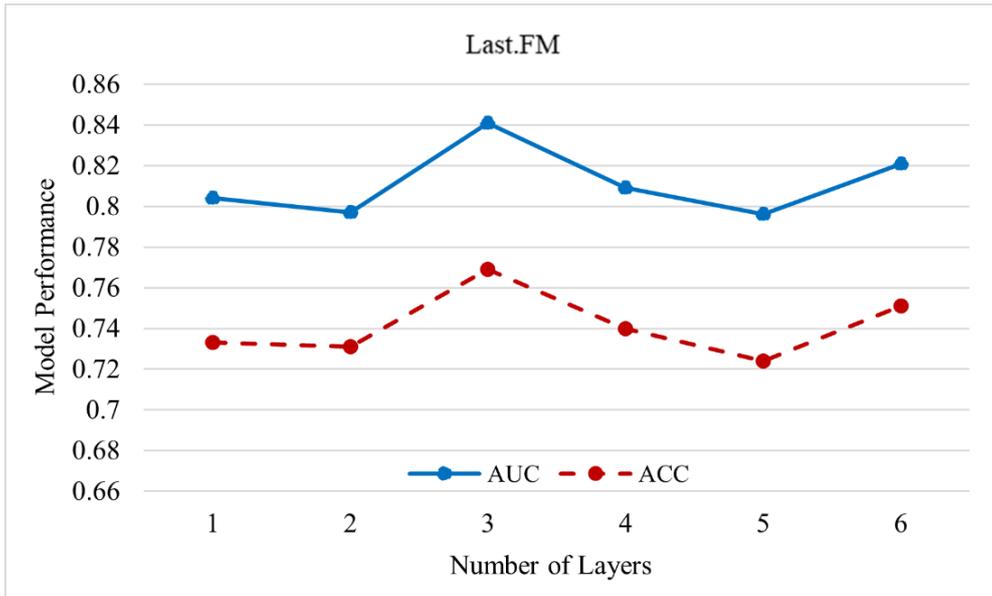

(c)

**Fig. 4** The influence of with different number of propagation layers on DEKGCI

*5.4.3 Influence of different aggregation method*

As mentioned above, three aggregation methods are used in the *aggregation layer* to yield the final user representation, namely, sum aggregation, concatenate aggregation, and neighbor aggregation. In this section, we carry out experiments to investigate the influence of different aggregation methods.

The AUC and ACC scores corresponding to different aggregation methods of the three datasets are provided in Fig.5(a) and Fig.5(b), respectively. It is clear that DEKGCI performs better with the sum aggregation than that of the concatenate aggregation and neighbor aggregation. One possible explanation is that the dimensionality of the vector does not increase during the sum aggregation processing, which makes the model less difficult to train.

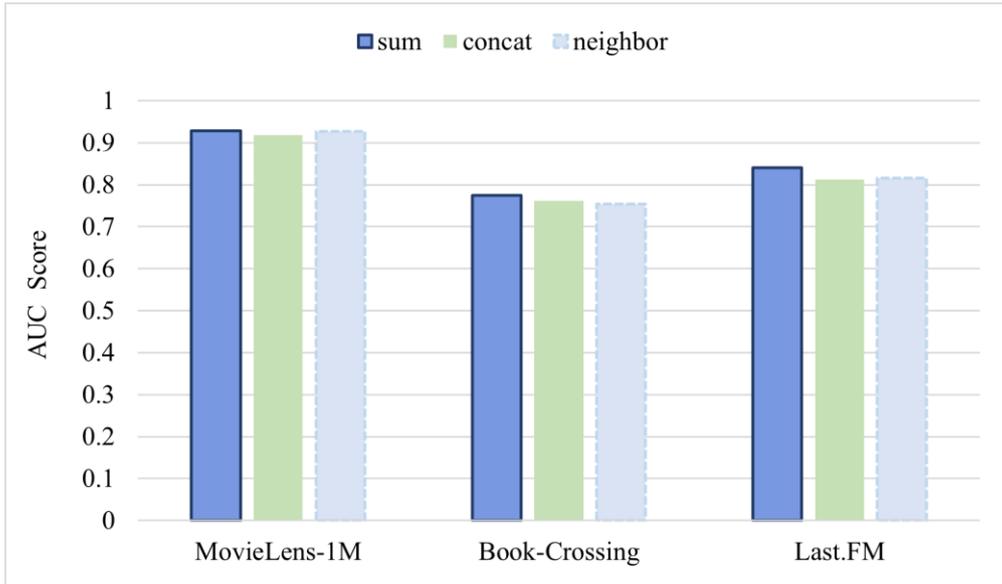

(a)

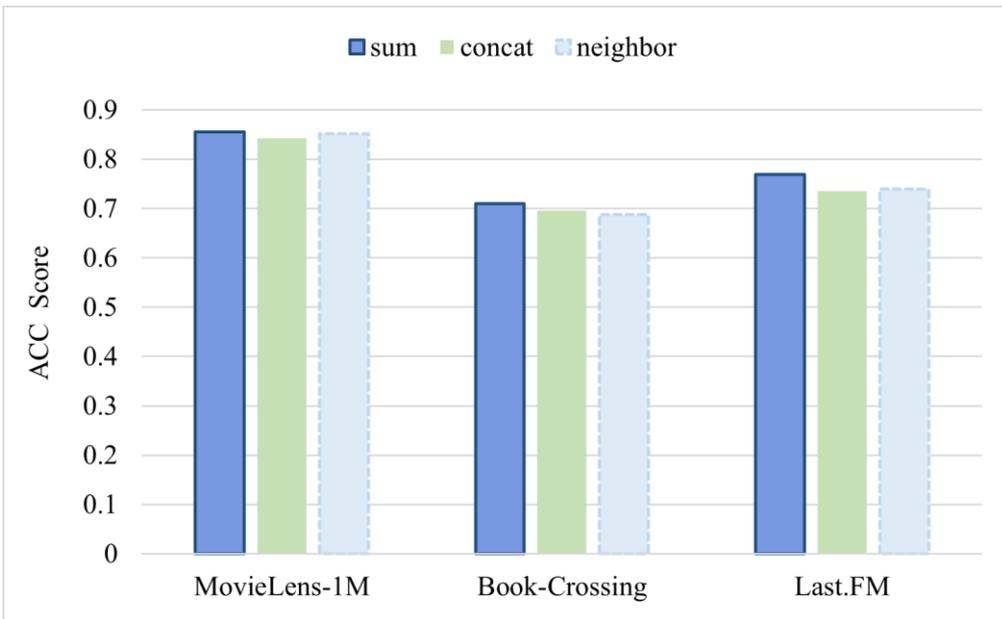

(b)

Fig. 5. The influence of different aggregators on DEKGCI

*5.4.4 Effect of the depth of receptive field*

The aggregation of multi-hop neighbor information on the KG can enrich the item representation to alleviate the data sparsity problem. In this section, we vary the depth of neighbor receptive field (i.e., the layers of covered neighbors) to investigate its influence on the performance of DEKGCI.

It can be seen from Fig.6(a)~(c) that the values of AUC and ACC decrease with the increase of the depth of receptive field for the three datasets. When the depth is 1, DEKGCI has the best performance. Therefore, only 1-order neighbors are considered for modeling the item in the proposed method. One possible reason is that noises are introduced when aggregating the long-distance neighbors' information.

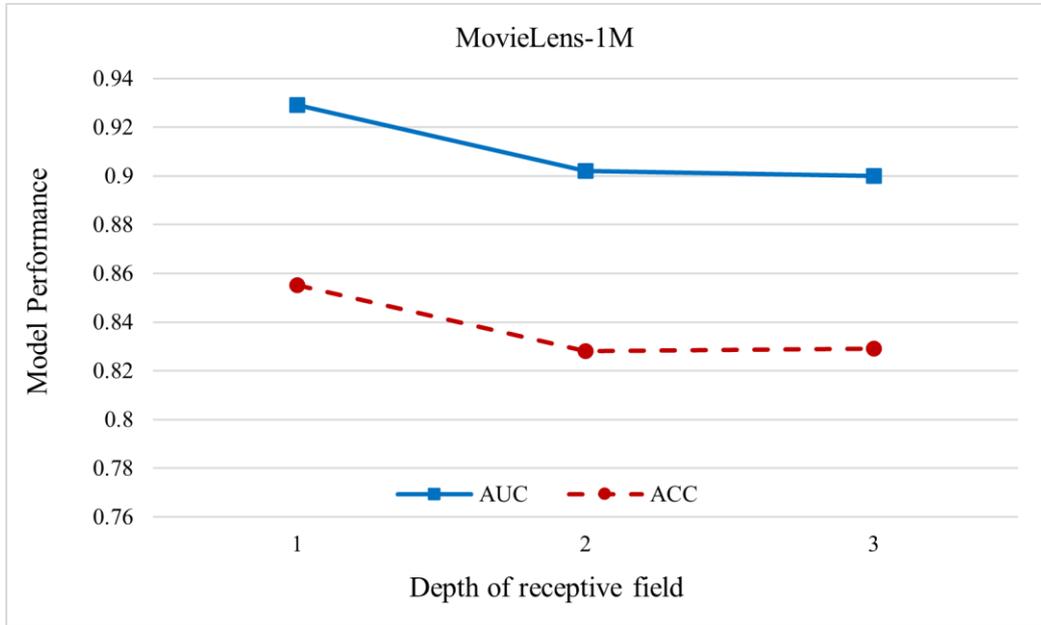

(a)

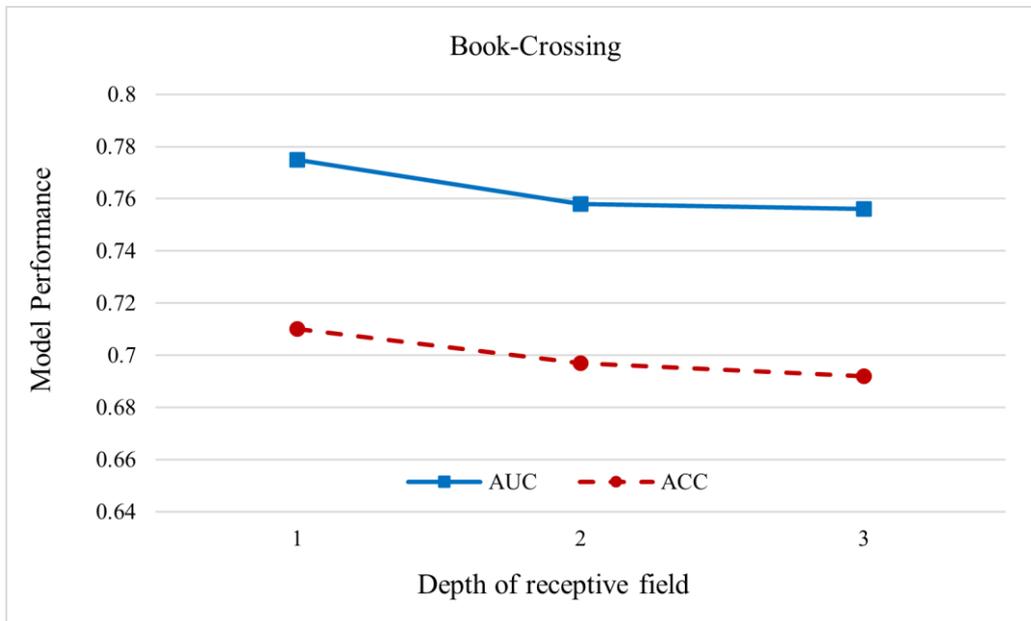

(b)

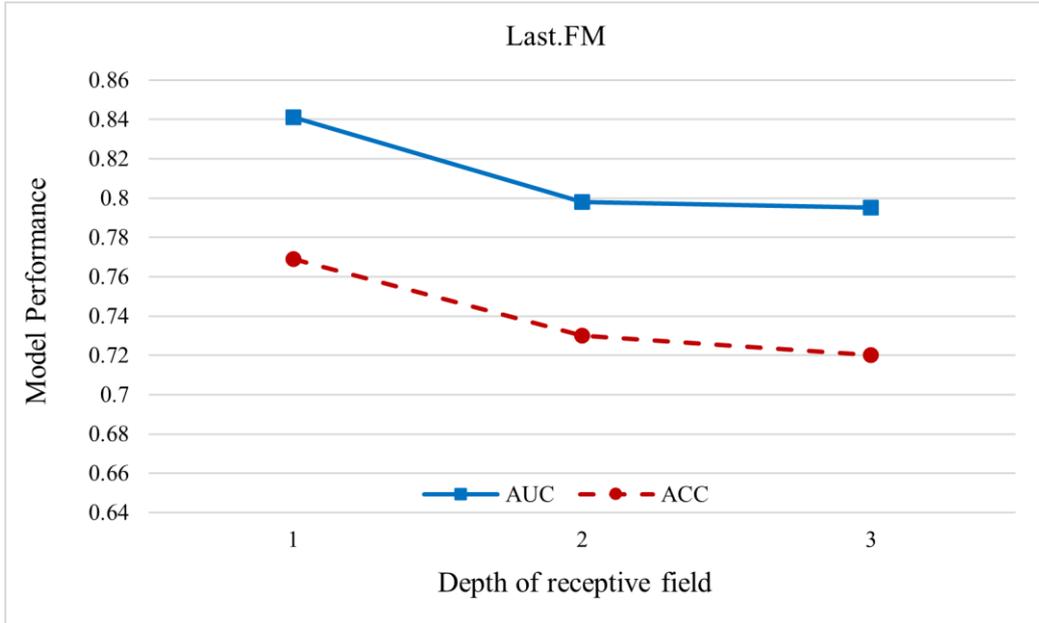

(c)

Fig. 6. The influence of different depth of receptive field on DEKGCI

Fig. 6(a)~(c) depicts how the depth of neighbor receptive field affects model performance. When a single layer of neighbor information is used to form the item representation, the model performance is best, as the usage of simple neighborhood information gathers the most pertinent neighborhood information and makes the model lighter.

6. Conclusion

Although extensive research has been conducted on the use of knowledge graphs and user-item interaction graphs in recommendation systems, there has been no attempt to combine both approaches in the same recommender model and learn both user and item representations simultaneously. Building on previous studies that utilized knowledge graphs or user-item interaction graphs in recommendation models, we propose DEKGCI, a double-sided recommendation model that optimizes both user and item representations concurrently. We conducted experiments on three commonly used datasets and found that our proposed algorithm outperforms seven state-of-the-art baselines in terms of AUC and ACC.

Our study confirms the value of integrating diverse sources of information to enhance user and item representations within a single recommendation model. For future research, we will explore the incorporation of other supporting information, such as social networks and user attributes, into the recommendation system. Additionally, we will investigate the potential application of our recommendation model in other domains, such as course recommendation.

## Acknowledgments

This work was financially supported by the Research Funds from the National Natural Science Foundation of China (Grant No. 62077019).

## Conflict of Interest

The authors declare that there does not exist any kind of conflict of interest.

## Data availability

The datasets utilized in this study can be accessed from the following websites: https://github.com/hwwang55/KGCN and https://github.com/hwwang55/RippleNet.